\documentclass[showpacs,prd,draft,
preprint,%tightenlines,
nofootinbib,12pt]{revtex4}

\usepackage{bm}
\usepackage{epsf}

\newcommand{\act}{{\cal S}}
\newcommand{\ord}{{\cal O}}
\newcommand{\be}{\begin{equation}}
\newcommand{\ee}{\end{equation}}
\newcommand{\bea}{\begin{eqnarray}}
\newcommand{\eea}{\end{eqnarray}}
\newcommand{\nl}{\nonumber \\}

\newcommand{\psibar}{\overline{\Psi}}
\newcommand{\del}{{\bf \Delta}}

\newcommand{\delv}{{\bf \nabla}}
\newcommand{\delvt}{\tilde{{\bf \nabla}}}

\newcommand{\Mbz}{{(aM_0)}}
\newcommand{\delfour}{{\Delta^{(4)}}}
\newcommand{\delsq}{\Delta^{(2)}}

\newcommand{\Ev}{\tilde{{\bf E}}}
\newcommand{\Bv}{\tilde{{\bf B}}}

\newcommand{\sigmav}{\mbox{\boldmath$\sigma$}}
\newcommand{\kp}{k^\prime}
\newcommand{\VEC}[1]{{\bf \bm{#1}}} % for bold 3-vectors

\begin{document}

%\draft
\title{{\bf One-Loop Matching of the Heavy-Light $A_0$ and $V_0$
       Currents with NRQCD Heavy and Improved Naive Light Quarks}}
\author{\large{Emel Gulez, Junko Shigemitsu, and Matthew Wingate}}
\altaffiliation{Present address: Institute for Nuclear Theory,
University of Washington, Seattle, WA 98115}
\affiliation{Physics Department, The Ohio State University,
  Columbus, OH 43210, USA}
%\date{DRAFT-- 9 December 2003.}

\begin{abstract}
One-loop matching of heavy-light currents is carried out for a 
highly improved lattice action, 
including the effects of mixings with dimension 4 $\ord(1/M)$ and
$\ord(a)$ operators. 
We use the NRQCD action for heavy quarks, the Asqtad improved 
naive action for light quarks, and the Symanzik improved
glue action.  As part of the matching procedure we also present results 
for the NRQCD self energy and for massless Asqtad quark 
 wavefunction renormalization  with improved glue.
\end{abstract}
\pacs{11.15.Ha, 12.38.Bx, 12.38.Gc, 13.25.Hw}
\maketitle

\section{Introduction}

The importance of approaching the chiral limit in a controlled way 
in lattice simulations of heavy-light systems has become increasingly 
evident in recent years. 
Progress in reducing statistical and discretization errors has 
resulted in chiral extrapolation uncertainties, 
together with operator matching errors, dominating the final total 
error in heavy-light meson decay constant and form factor calculations.  
This has motivated the HPQCD collaboration to initiate studies
 of heavy-light systems using improved staggered/naive 
light quarks, taking advantage of the good chiral properties of
such actions which allow one
to go down to much smaller quark masses than has been possible in the 
past \cite{hlstagg,fbs,ourlat03,fermilat03}.  
Heavy-light simulations are now being carried out with light quark masses 
as low as $m_{strange}/8$, and contact has been established with chiral 
perturbation theory predictions.

\vspace{.2in}
\noindent
An important ingredient in all heavy meson decay constant and 
form factor calculations is the matching of the heavy-light currents 
used in the simulations to their continuum QCD counterparts.  
 The highly improved  heavy-light actions introduced above necessitate a new 
round of matching calculations.
 In this 
article we report on the one-loop perturbative matching of the 
temporal component of the heavy-light axial current, $A_0$,  with 
NRQCD heavy quarks and massless Asqtad naive quarks. These matching 
coefficients are directly relevant for our ongoing $f_B$, $f_{B_s}$ and 
$f_{D_s}$ calculations on the MILC dynamical configurations and 
have already been applied in the results quoted in \cite{hlstagg,fbs,ourlat03}.
Due to the chiral symmetry of naive quarks, matching coefficients 
for the vector current are identical to those of the axial current;
thus the results presented here can also be applied to form factor 
calculations \cite{shigelat03}.  
Our matching calculations include contributions from 
$1/M$ current corrections and  an $\ord(a \; \alpha_s)$ 
discretization correction. The matched heavy-light current 
is then correct through $\ord(\alpha_s)$, $\ord(a  \; 
\alpha_s)$, $\ord(\alpha_s/(aM))$ and $\ord(\alpha_s \, \Lambda_{QCD}/M)$.  
Further corrections would come in at $\ord(\alpha_s^2)$,
$\ord(\Lambda^2_{QCD}/M^2)$ and $\ord(a^2  
\; \alpha_s)$.

\vspace{.2in}
\noindent
In the next section we list the quark and glue actions employed and
 give a brief discussion of our calculational methods. Two 
independent strategies were adopted and the results tested against 
each other.  The wavefunction renormalization 
constant $Z_q$ for massless Asqtad fermions, which is used in later sections
on matching coefficients, is presented in section 3. 
Section 4 describes one-loop self-energy corrections for NRQCD 
heavy quarks.  Both the heavy quark wavefunction renormalization 
$Z_Q$ and the mass renormalization $Z_M$ enter into the 
current matchings.  Section 5 presents the full mixing and matching 
calculations for the NRQCD/Asqtad heavy-light currents. We give tables 
of results for a range of heavy quark masses. 
We add several appendices with calculational details, such as a list 
of Feynman rules.

\section{The Lattice Actions and Calculational Strategies}

\subsection{ The Glue Action}

The tree-level tadpole and $\ord(a^2)$ improved glue action 
is given by \cite{weisz},
\be
\label{glueaction}
\act_G = - \beta \sum_{x,\,\mu > \nu}  \left\{
\frac{5}{3} \frac{P_{\mu \nu}}{u_0^4} 
- \frac{1}{12} \frac{R_{\mu \nu}}{u_0^6} 
- \frac{1}{12} \frac{R_{\nu \mu}}{u_0^6} \right\} ,
\ee
with
\begin{eqnarray}
P_{\mu \nu} &=& \frac{1}{N_c}  {\rm Re} \left( {\rm Tr} \{ U_\mu(x)
 U_\nu(x+a_\mu) U^\dagger_\mu(x+a_\nu) U^\dagger_\nu(x) \} \right) ,  \\
R_{\mu \nu} &=& \frac{1}{N_c}  {\rm Re} \left( {\rm Tr} \{ U_\mu(x)
 U_\mu(x+a_\mu) U_\nu(x+2a_\mu) 
U^\dagger_\mu(x+a_\mu + a_\nu) U^\dagger_\mu(x+a_\nu) U^\dagger_\nu(x) \}
 \right)  .
\end{eqnarray}
$\beta = 2N_c/g^2$ and 
$u_0$ is the tadpole improvement factor
 \cite{lepmac}.  
We carry out perturbative calculations in general covariant gauge and add a
gauge fixing term,
\be
\label{sgfa}
\act_{gf} =
 \frac{1}{2 \xi} \,  \sum_x \left[\sum_\mu
 \partial_\mu (a \, A_\mu) \right]^2 ,
\ee
with $\partial_\mu A_\mu(x) \equiv A_\mu(x + a_\mu/2) - 
A_\mu(x - a_\mu/2)$ and $\xi$ the gauge parameter. We work in both 
the Feynman and the Landau gauges and verify that gauge invariant quantities 
are independent of $\xi$. 
In one-loop matching calculations of current operators, 
the only ingredient required from the gauge action is the tree-level
gauge propagator. The form of the free gauge propagator for 
improved gauge actions has been discussed in several articles \cite{weisz,aoki,
groote}.
 We follow reference \cite{groote} and invert the momentum space 
 free improved gauge action 
 once and for all using Mathematica.
One ends up with a closed form analytic expression
for the gauge propagator which is a $4 \times 4$ matrix in 
Lorentz space.  The interested reader is 
referred to Appendix A of \cite{groote} for more details.

\subsection{ The Light Quark Action}

The ``Asqtad'' action is the most successful quark action 
to date for simulating  light dynamical and valence quarks \cite{naik,peter,
doug}. 
The MILC and 
 HPQCD collaborations are using the MILC dynamical configurations 
to study light-light, heavy-light and heavy-heavy physics 
\cite{milc,claude,agray,prl,gottlieb}\cite{fbs,ourlat03,fermilat03}. 
 This action 
has been designed to drastically reduce the taste-symmetry breaking 
effects of conventional staggered fermions and also has all 
other $\ord(a^2)$ discretization errors removed.  
As discussed in references \cite{peter} \cite{hlstagg}
 there are two equivalent 
ways to write down the Asqtad action, either in terms of four component 
``naive'' fermions or in terms of one component ``staggered'' fields.  
We prefer to use the naive fermion approach. We will use Feynman rules 
for four component fields and our heavy-light currents will be 
constructed out of improved naive light quark fields.

\vspace{.2in}
\noindent
Three new ingredients are necessary to go from the unimproved 
staggered/naive action to the Asqtad action, fattening of links, 
the so-called ``Lepage correction'' and the Naik term. One starts from 
the unimproved naive action,
\be
\label{nveaction}
{\cal S}_0 = a^4 \sum_x \left\{ \psibar (x) \,\left[ \sum_\mu \gamma_\mu
{\frac{1}{a}}\nabla_\mu \; + \; m \right] \Psi(x) \right\},
\ee
with
\be
\label{nabla}
\nabla_\mu \Psi(x) = \frac{1}{2 \, u_0} \,[ U_\mu(x) \Psi(x+a_\mu) - 
 U^\dagger_\mu (x-a_\mu)\Psi(x-a_\mu)] \, ,
\ee
and Hermitian Euclidean $\gamma$-matrices obeying
$\{\gamma_\mu,\gamma_\nu \} = 2 \delta_{\mu \nu}$.
The link fattening and the Lepage terms are incorporated by replacing 
the link variable $U_\mu$ in the covariant derivative by an updated 
variable $V^\prime_\mu$ (the corresponding covariant derivative will be 
denoted $\nabla^\prime_\mu$).
\be
\label{eq:vprime}
V_\mu'(x) ~\equiv~ V_\mu(x) - \sum_{\rho\ne\mu} \frac{(\nabla^{\ell}_\rho)^2}
{4} U_\mu(x) 
\ee
\be
\label{vmu}
V_\mu(x) ~\equiv~ \prod_{\rho\ne\mu} \left(1 + \frac{\nabla^{\ell,(2)}_\rho}{4}
\right)\bigg|_{\rm symmetrized} U_\mu(x)
\label{eq:fatlink}
\ee
$V_\mu(x)$ is the fattened (taste violation suppressing) link 
and the second term in eq.(\ref{eq:vprime}) is the Lepage term that removes 
a low momentum $\ord(a^2)$ error.  The derivatives, $\nabla^{\ell}_\nu$
 and $\nabla^{\ell,(2)}_\nu$, that act on link matrix variables are defined as
 ($\mu \ne \nu$),
\begin{eqnarray}
\label{linkdel}
\frac{1}{u_0}\nabla^{\ell}_\mu \,U_\nu(x) &=& \frac{1}{2 \, u_0^3}
\,\Big[ U_\mu(x) \, U_\nu(x+a_\mu) \, U^\dagger_\mu(x+a_\nu) \nl
&& \;-\;
U^\dagger_\mu(x-a_\mu) \, U_\nu(x-a_\mu) \, U_\mu(x-a_\mu+a_\nu) \Big] \, ,
\\
\frac{1}{u_0}\nabla^{\ell,(2)}_\mu\, U_\nu(x) &=& \frac{1}{u_0^3 }
\,\Big[ U_\mu(x) \, U_\nu(x+a_\mu) \, U^\dagger_\mu(x+a_\nu) \nl
&& \;+\;
U^\dagger_\mu(x-a_\mu) \, U_\nu(x-a_\mu) \, U_\mu(x-a_\mu+a_\nu) \Big]
\nl && - \frac{2}{u_0} \, U_\nu(x) \, .
\end{eqnarray}
The Naik correction adds a three-link term to the modified covariant 
derivative so that the final form of the Asqtad action takes on the form
\be
\label{asqtadact}
{\cal S}_{Asqtad} = a^4 \sum_x \left\{ \psibar (x) \,\left[ \sum_\mu \gamma_\mu
{\frac{1}{a}} \left( \nabla^\prime_\mu 
 \,- \, \frac{1}{6} \nabla_\mu^{3-link} \right) 
\; + \; m \right] \Psi(x) \right\} \, ,
\ee
with
\begin{eqnarray}
\nabla^{3-link}_\mu \Psi(x) &=& (\nabla_\mu)^3 \bigg|_{\rm tadpole \; improved}
\Psi(x) \nonumber \\
 &=& \frac{1}{8} \left\{ \frac{1}{u_0^3} \, \left[ 
UUU \,\Psi(x + 3 a_\mu) - U^\dagger U^\dagger U^\dagger \,\Psi(x - 3 a_\mu) 
\right] \right.  \nonumber \\
  & & \quad \left. - \frac{3}{u_0} \left[U \,\Psi(x + a_\mu) - U^\dagger 
\,\Psi(x - a_\mu \right] \right\} \, .
\end{eqnarray}
Feynman rules for the action (\ref{asqtadact}) have been written 
down in concise form by Q. Mason \cite{quentin}.
  We verified the rules for the one-gluon emission vertices 
 and for the subset of two-gluon emission vertices relevant to the 
present calculations, namely the subset symmetric in the two gluons. 
 These Feynman 
rules are summarized in the appendix.

\subsection{ The Heavy Quark Action}

We use the NRQCD action improved through $\ord(1/M^2)$ and $\ord(a^2)$, 
and which also includes the leading relativistic $\ord(1/M^3)$ correction
 \cite{cornell}. 
This action is currently being used in simulations of heavy-heavy and 
heavy-light systems on the MILC dynamical 
configurations. It has been discussed in 
many previous publications, and hence we will be brief here. In terms of 
the two-component Pauli spinor $\phi$ one has,
\bea
 \label{nrqcdact}
\act_{\rm NRQCD}  =
\sum_x \Bigg\{  {\phi}^\dagger_t \phi_t &-&
 {\phi}^\dagger_t
\left(1 \!-\!\frac{a \delta H}{2}\right)_t
 \left(1\!-\!\frac{aH_0}{2n}\right)^{n}_t \nonumber \\
& \times & \frac{1}{u_0} \,
 U^\dagger_t(t-1)
 \left(1\!-\!\frac{aH_0}{2n}\right)^{n}_{t-1}
\left(1\!-\!\frac{a\delta H}{2}\right)_{t-1} \phi_{t-1} \Bigg\} \, .
 \eea
 $H_0$ is the nonrelativistic kinetic energy operator,
 \be
a H_0 = - {\delsq\over2\Mbz} \, ,
 \ee
and $\delta H$ includes relativistic and finite-lattice-spacing
corrections,
 \begin{eqnarray}
a\delta H
&=& - c_1\,\frac{(\delsq)^2}{8\Mbz^3}
+ c_2\,\frac{i}{8\Mbz^2}\left(\delv\cdot\Ev - \Ev\cdot\delv\right) \nl
& &
 - c_3\,\frac{1}{8\Mbz^2} \sigmav\cdot(\delvt\times\Ev - \Ev\times\delvt)\nl
& & - c_4\,\frac{1}{2\Mbz}\,\sigmav\cdot\Bv
  + c_5\,\frac{\delfour}{24\Mbz}  - c_6\,\frac{(\delsq)^2}
{16n\Mbz^2} \, .
\label{deltaH}
\end{eqnarray}
All derivatives are tadpole improved and,
\be
\delsq = \sum_{j=1}^3\nabla_j^{(2)}, \qquad \qquad \delfour = \sum_{j=1}^3
\nabla_j^{(4)}
\ee
\be
\tilde{\nabla}_k = \nabla_k - \frac{1}{6}\nabla_k^{(3)} \, .
\ee
Precise definitions of the $\nabla_j^{(i)}$, $ i = 2,3,4$, and 
of the improved $\Ev$ and $\Bv$ 
field operators are given, for instance,
 most recently in Appendix B of \cite{hlstagg} (we note, however, a 
factor of 2 error in eq.(B4) of this reference, which is corrected 
above in eq.(\ref{linkdel})).
Feynman rules for simpler versions of NRQCD actions have appeared 
in \cite{bethch,pert2,hashimoto}.
  We have generalized them to rules for the more 
highly improved action
 (\ref{nrqcdact}) - (\ref{deltaH}) studied here.
 The new Feynman rules are given in an appendix.

\subsection{ Calculational Strategies }
We have employed two independent approaches to carrying out the 
one-loop matching calculations with the above actions and used 
them as checks against each other.  In one method we make extensive use 
of Mathematica to multiply propagators and vertices, to carry out the 
Dirac algebra, and, where necessary, to take derivatives with respect 
to external momenta.  The resulting expressions are then converted 
into FORTRAN and fed into a FORTRAN VEGAS code for the numerical 
integration over internal momenta.  

\vspace{.2in}
\noindent
In many instances the Mathematica 
expressions can be kept quite simple and general,
 since details of Feynman rules are 
only needed at the VEGAS stage. For instance, in calculations  that 
do not involve any derivatives with respect to external momenta, 
the only information the Mathematica code requires is the formal structure 
of the vertices in the Asqtad and NRQCD actions.  For the emission of a 
gluon with polarization $\mu$, the Asqtad action has the general 
vertex,
\be
\label{vq}
V(A_\mu) = \sum_\nu w_{\mu , \nu} \; \gamma_\nu
\ee
and the NRQCD action leads to 
\be
\label{vh}
V_H(A_\mu) =   wh_{\mu,0} + \sum_j wh_{\mu , j}\; \sigma_j 
\ee
 $\sigma_j$ = Pauli matrices (see end of Appendix A.1 for a 4 component 
version of (\ref{vh})).
The dependence of the ``$w$'''s  on the incoming and outgoing fermion 
momenta is specified only in the VEGAS code through subroutines that 
code up the Feynman rules. 

\vspace{.2in}
\noindent
Mathematica is also very useful when one needs to take
derivatives with respect to external momenta.  One inserts the Feynman rule 
expression for the $w_{\mu , \nu}$ or $wh_{\mu , \nu}$ of interest and 
 lets Mathematica take derivatives before setting 
external momenta to zero.

\vspace{.2in}
\noindent
Our second method is based on a $C^{++}$ code developed by C.\ Morningstar 
for matching calculations with NRQCD/clover quark actions and which was 
generously made available to us. We modified the original code 
to replace clover by the Asqtad light quark action and also to 
accommodate
 improved glue with non-diagonal (in Lorentz space) gauge propagators.
The $C^{++}$ code handles derivatives via ``automatic differentiation.''
A $C^{++}$ class is defined which carries information 
not just about a function, but also about its first couple of 
derivatives through a Taylor series expansion.  When two 
such class instances are multiplied, for example, 
derivatives of the product are calculated 
automatically via the chain rule as part of the definition of an 
overloaded multiplication operation.  
We believe our two methods are sufficiently independent of each other 
so that they provide good checks of our results.

\section{$Z_q$ for Massless Asqtad Quarks}
Perturbative 
calculations with the one-component (improved staggered) 
version of the Asqtad action and with unimproved Wilson glue have been 
performed recently in \cite{leesh} for the matching of 
light-light bilinear and four-fermion operators.  Calculations using 
the four-component (improved naive fermion)
 Asqtad action and improved glue 
have been carried out for bilinear and 
four-fermion operators \cite{trottier1} 
 and for mass renormalization \cite{jhein}.
For the matching of heavy-light currents the light quark wavefunction 
renormalization $Z_q$ is required.  This quantity has been 
derived already with improved glue using ``twisted'' boundary conditions
\cite{trottier2}.
 We have repeated the calculation with a gluon mass IR regulator, 
 the IR regulator we use throughout this article. 
 $Z_q$ is relevant for many other light-light and heavy-light 
perturbative calculations.  We list our results here for the
case of massless light quarks.
\be
\label{zq}
Z_q = 1 + \alpha_s \, [\,C^{IR}_q  \; + \; C_q\,] \; + \; \ord(\alpha_s^2)
\ee
\be
C^{IR}_q = \frac{1}{3 \, \pi} \left[1 + (\xi - 1) \right] \; 
{\rm ln}( a^2 \lambda^2)
\ee
where $\lambda$ is the gluon mass. The gluon mass dependence cancels, 
of course, 
upon taking the difference between continuum and lattice one-loop 
coefficients. The IR finite $C_q$ has contributions from the regular 
rainbow diagram, the tadpole diagram, and from the $u_0$ tadpole improvement 
procedure.  We write
\be 
C_q = C_q^{reg} \; + \; C_q^{tad} \; + \; C_q^{u0}
\ee
and list results in Table I for both Feynman and Landau gauges. The 
$u_0$ correction contribution is given by 
\be
\label{cqu0}
C_q^{u0} = \; - \; \left[\,4 - \frac{1}{4} - \frac{3}{2}\right] \,
 u^{(2)}_0 = \; - \; \frac{9}{4} \; u^{(2)}_0 \, ,
\ee
with $u^{(2)}_0$ defined through,
\be
u_0 \equiv 1 - \alpha_s \, u^{(2)}_0 + \ord(\alpha_s^2) \, .
\ee
The three contributions in (\ref{cqu0}) come from the fat-link, 
Lepage,  and Naik terms respectively.  The numerical value of $u_0^{(2)}$ 
depends on how one defines $u_0$ and on the gauge action.  In the case of 
improved glue one has $u_0^{(2)} = 0.767$ for 
the plaquette definition of $u_0$ and $u_0^{(2)} = 0.750$ for 
the Landau link definition.  The MILC dynamical configurations 
were created using the plaquette $u_0$. It then makes sense to use 
plaquette $u_0$ in nonperturbative valence light quark propagators and 
 in perturbative light quark self energy calculations.
For the NRQCD heavy quark parts of heavy-light simulations 
 we have adopted the Landau link $u_0$ and this is reflected in the 
heavy quark parts of the perturbative matching given below.  
  There is no contradiction in using different definitions of $u_0$ 
in different parts of a calculation. The important thing is that 
the perturbation theory match the choices made in the numerical 
simulations.  We give explicit formulas for the tadpole improvement 
contributions (such as (\ref{cqu0}) and several others below) so readers
should be able to adjust the numbers in our Tables to their $u_0$ choices. 

\section{Heavy Quark Self Energy}

\begin{figure}[t]
\epsfxsize=\hsize
\epsfbox{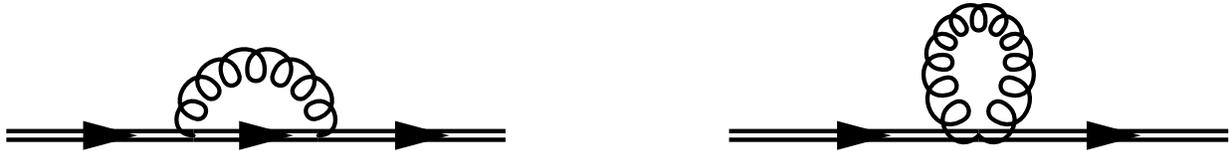}
\caption{\label{fig:SEdiag}
Self-energy diagrams % (labeled (a)--(b) in the text)
which generate $C_Q$ (or $C_q$ if a light quark line is substituted).}
\end{figure}

One-loop self-energy calculations have appeared already in many articles 
 for a variety of NRQCD actions on the lattice.  
Refs.~\cite{bethch,hashimoto} present results for simpler NRQCD 
actions with unimproved glue, Ref.~\cite{colin}
 deals with highly improved NRQCD actions 
and Wilson glue, and Ref.~\cite{jones} works with Symanzik improved glue but at
low order in NRQCD.  Here we summarize results for the NRQCD action of 
eqns.\ (\ref{nrqcdact}) - (\ref{deltaH}) and improved glue.  We are 
interested in wavefunction renormalization $Z_Q$, mass renormalization 
$Z_M$, and (for completeness) the energy shift $E_0$.
\bea
\label{zhq}
Z_Q &=& 1 \; + \; \alpha_s \, [\, C_Q^{IR} \; + \; C_Q \,] \; + \;
\ord(\alpha_s^2)  \\
Z_M &=& 1 \; + \; \alpha_s \,  C_M   \; + \;
\ord(\alpha_s^2)  \\
a \, E_0 &=& \qquad \; \alpha_s \,  C_{E0}   \; + \;
 \ord(\alpha_s^2) 
\eea
where,
\be
C^{IR}_Q = \frac{1}{3 \, \pi} \left[-2 + (\xi - 1) \right] \; 
{\rm ln}( a^2 \lambda^2) \, .
\ee
Just as for light quarks, the one-loop heavy quark self energy 
$ \alpha_s \,{\bf \Sigma (p)} $
gets contributions from the two diagrams in Fig.~1 and from tadpole 
improvement.  If one defines
\bea
\label{omegas}
\Omega_0 &=& \;\;\;{\bf \Sigma (0)} \\
\label{omega1}
\Omega_1 &=& - \, i \, \frac{\partial {\bf \Sigma}}{\partial p_0} 
\bigg|_{p_\mu = 0}  \\
\Omega_2 &=& 2(aM) \, 
\frac{\partial {\bf \Sigma}}{\partial \VEC{p}^2} 
\bigg|_{p_\mu = 0}  = (aM) \,
\frac{\partial^2 {\bf \Sigma}}{\partial p_x \, \partial p_x} 
\bigg|_{p_\mu = 0}  
\eea
where $p_\mu$ stands for 
 the heavy quark external momentum measured 
in units of the inverse lattice spacing (${\bf \Sigma }$ is also 
taken to be dimensionless) 
and $M$ is the heavy quark pole mass,
then,
\bea
\label{cxx}
C_{E0} &=&  - \; \Omega_0 \\
C_Q &=& \;\Omega_0 + \Omega_1  \bigg|_{{\rm IR \; finite}} \\
C_M &=& \; \Omega_2 - \Omega_1 \, .
\eea
Results for the one-loop coefficients $C_{XX}$ are tabulated in Table II 
for several values of $(aM_0)$ and $n$. 
To the order we are working there is no need to distinguish between the pole 
mass $M$ and the bare mass $M_0$ in one-loop coefficients. 
 Since simulations are carried out at 
fixed $(aM_0)$ it is much more convenient to express everything 
in terms of the bare mass. 
 The results in Table II were obtained using the Landau link $u_0$, 
with the explicit formulas given by,
\bea
C_{E0}^{u0} &=&   \left[\, - \,1 \, -  \, \frac{3}{(aM_0)} \,-\,
 \frac{c_5}{2 (aM_0)} 
\,+\, \frac{3}{2} \left(\frac{c_1}{(aM_0)^3} + \frac{c_6}{2n(aM_0)^2}
\right) \, \right] \; u^{(2)}_0  \\
C_M^{u0} &=&  \left[ \,-\,1  \,+\, \frac{3}{2n(aM_0)} \,+\, \frac{c_5}{3} 
\,-\, 3\,(aM_0)\, \left(\frac{c_1}{(aM_0)^3} + \frac{c_6}{2n(aM_0)^2}
\right) \, \right] \; u^{(2)}_0 \;.
\eea
There are no tadpole improvement corrections to $C_Q$ since such 
corrections cancel between $\Omega_0$ and $\Omega_1$, as do the 
contributions from the tadpole diagram.
Wherever possible we give analytic formulas as functions of the 
coefficients $c_i$ in the NRQCD action eq.(\ref{deltaH}), however 
numerical results such as in Table II are always given with  $c_i=1$.

\vspace{.1in}
\noindent
$Z_M$ and $E_0$ are gauge invariant, however $Z_Q$ is both gauge variant 
and IR divergent.  We discuss in an appendix how we handle IR
divergent terms such as in (\ref{zhq}) and isolate the IR finite 
contribution.

\section{Matching of the Heavy-Light Current}
The basic formalism for perturbative matching of heavy-light currents 
has been developed already for NRQCD/clover currents \cite{pert2} and 
can be taken over without any modification for the case of improved 
naive light quarks. Again we will be brief and refer the reader 
to the earlier articles for details. 
The following three currents are needed in the NRQCD effective theory 
  to match the temporal component 
of the vector or axial vector currents 
to full QCD through $\ord(\alpha_s/M)$.
\begin{eqnarray}
\label{j0}
 J^{(0)}_{0}(x) & = & \bar q(x) \,\Gamma_0\, Q(x),  \\
\label{j1}
 J^{(1)}_{0}(x) & = & \frac{-1}{2 \,(aM_0)} \bar q(x)
    \,\Gamma_0\,\mbox{\boldmath$\gamma\!\cdot\!\nabla$} \, Q(x), \\
\label{j2}
 J^{(2)}_{0}(x) & = & \frac{-1}{2 \,(aM_0)}  \bar q(x)
    \,\mbox{\boldmath$\gamma\!\cdot\!\overleftarrow{\nabla}$}
    \,\gamma_0\ \Gamma_0\, Q(x). 
\end{eqnarray}
The $Q$ fields are four component Dirac spinors with upper two 
components given by the NRQCD $\phi$ field and vanishing lower 
components, and 
$\Gamma_0 = \gamma_0$ or $\gamma_5  \gamma_0$.  The $\nabla$ in 
$J^{(1)}_0$ and $J^{(2)}_0$ is the same as in (\ref{nabla}).
The relation 
between matrix element of $A_0$ in full QCD 
to those of the effective theory currents can be 
written as,
\begin{eqnarray}
\label{a0}
\langle \, A_0 \, \rangle  &=& ( 1 + \alpha_s \, 
\tilde{\rho}_0)\,\langle J^{(0)}_0 \rangle + \nonumber \\
 & & (1 + \alpha_s   \,  \rho_1) \, \langle 
J^{(1),sub}_0 \rangle  \nl
 & & \quad  \; + \; \alpha_s  \,
 \rho_2 \, \langle J^{(2),sub}_0 \rangle 
\quad + \quad \ord(\alpha_s^2, \Lambda_{QCD}^2/M^2, a^2 \, \alpha_s)  \, ,
\end{eqnarray}
with
\be
\label{jsub}
 J_0^{(i),sub} = J_0^{(i)} - \alpha_s \,   \zeta_{i0}  \,
J_0^{(0)} 
\ee
for $i = 1,2$. 
We prefer to express things in terms of the
more physical   matrix elements $\langle J_0^{(i),sub} \rangle$ 
which have had $\ord(\alpha_s/(aM_0))$ power law contributions 
subtracted out \cite{fbscale}.
  The coefficients $\tilde{\rho}_0$, $\rho_1$ and 
$\rho_2$ are then given by,
\bea
\label{rhos}
\tilde{\rho}_0 & = & B_0 - \frac{1}{2} (C_q + C_Q) - \zeta_{00} \nl
\rho_1 & = & B_1 - \frac{1}{2} (C_q + C_Q) - C_M - 
\zeta_{01} - \zeta_{11} \nl
\rho_2 & = & B_2 - \zeta_{02} - \zeta_{12}
\eea
with
\bea
\label{bfinite}
B_0 &=& \frac{1}{\pi} \left[ {\rm ln}(a M_0) - \frac{1}{4} \right] \nl
B_1 &=& \frac{1}{\pi} \left[ {\rm ln}(a M_0) - \frac{19}{12} \right] \nl
B_2 & = & \frac{4}{\pi}
\eea
$C_q$, $C_Q$ and $C_M$ are the one-loop coefficients of self-energy 
corrections discussed in the two previous sections. $C_M$ enters into 
$\rho_1$ because we have written the $1/M$ currents in (\ref{j1}) 
and (\ref{j2}) 
using $M_0$ rather than the pole mass.  Originally these currents 
are defined in terms of the pole mass, the  mass 
that is common to both full QCD and the effective lattice 
theory. One then, for convenience, replaces $M$ by $Z_M  M_0$ and 
notes that to one-loop order one can set $Z_M = 1$ everywhere 
except in the tree-level contribution from $J_0^{(1)}$, where 
$Z_M = 1 + \alpha_s \, C_M$ must be used.

\begin{figure}
\epsfxsize=\hsize
\epsfbox{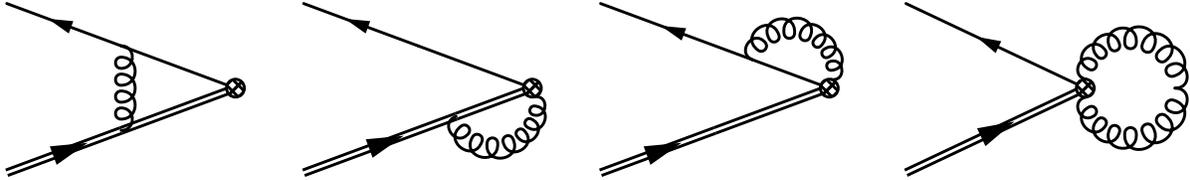}
\caption{\label{fig:HLdiag}
One loop vertex corrections which generate the $\zeta_{ij}$.}
\end{figure}

\vspace{.1in}
\noindent
The $\zeta_{ij}$ in (\ref{jsub}) and (\ref{rhos}) 
come from the mixing between the three currents. 
They are calculated at one-loop from diagrams shown in Fig.~2 
where one puts $J_0^{(i)}$ at the vertex and projects out the 
tree-level expression  $\langle J_0^{(j)} \rangle_{tree}$.  
For instance, in the case of the $\zeta_{i0}$ that 
appear in the power law subtraction (\ref{jsub}), one is taking the 
matrix element of $J_0^{(i)}$ and asking how much of it projects back onto 
$J_0^{(0)}$.
 Since we match at zero external momentum
one has $\zeta_{10} = \zeta_{20}$. As explained in 
references \cite{pert2,shig} $\rho_2$, 
 or more specifically $\zeta_{02}$, includes 
a term that removes an $\ord(a \, \alpha_s)$ discretization error 
from $J_0^{(0)}$.  The matching procedure is such that $\ord(\alpha_s/M)$ 
and $\ord(a \, \alpha_s)$ corrections are (and must be) made at 
the same time.

\vspace{.1in}
\noindent
We mention that some care is required in calculating $\zeta_{i2}$ , 
 $i = 0,1$.  There is another dimension 4 current operator 
involving a time derivative acting on the light quark field 
$\bar q(x)$ which is equivalent to $J_0^{(2)}$ via equations of motion. 
\be
 \tilde{J}^{(2)}_{0}(x)  =  \frac{1}{2 \, (aM_0)} \,  \bar q(x)
   \overleftarrow{\nabla}_0
    \, \Gamma_0\, Q(x). 
\ee
When calculating $\zeta_{i2}$ one must also include contributions from 
  $\langle \tilde{J}_0^{(2)} \rangle_{tree}$.  

\vspace{.1in}
\begin{figure}[t]
\epsfxsize=0.4\hsize
\hspace{.1mm}
\epsfbox{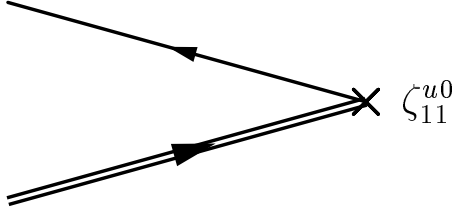}
\caption{\label{fig:HLmlink}
$u_0$ correction to $\rho_1$: $\zeta_{11}^{u0}$.}
\end{figure}
\noindent
Our results for $\tilde{\rho}_0$, $\rho_1$, $\rho_2$ and $\zeta_{10}$ 
are given in Table III.  
Only $\rho_1$ has a $u_0$ correction, coming from the contribution
 to $\zeta_{11}$ from Fig.~3:
\be
\zeta_{11}^{u0} = u_0^{(2)} \, .
\ee
We use the Landau link $u_0$ in calculating $\rho_1$.  Table III
can be used in several ways. 
Should one choose not to include 
matrix elements of the $1/M$ currents $J_0^{(1)}$ and $J_0^{(2)}$ in 
one's simulation, then matching should be done with just the 
first term in eq.(\ref{a0}). In other words, one should not 
include the $- \, \alpha_s \,\zeta_{10} \, \langle J_0^{(0)} \rangle$ 
 term which we have absorbed into $\langle J_0^{(1),sub} \rangle$.  This 
way one avoids unnecessarily introducing  a lattice artifact 
$\ord(\alpha_s /(aM_0))$ power law term that can only be canceled by 
the matrix element $\langle J_0^{(1)} \rangle$. 
Using just the first term in (\ref{a0}) leaves us with 
$\ord(\Lambda_{QCD}/M)$ and $\ord(a \, \alpha_s)$ errors.  The next 
step in improvement would be to use 
 $\langle \, A_0 \, \rangle  = ( 1 + \alpha_s \, 
\tilde{\rho}_0)\,\langle J^{(0)}_0 \rangle \; + 
\; \langle J^{(1),sub}_0 \rangle  $.  Again there are 
no $\ord(\alpha_s /(aM_0))$ power law contributions and errors 
from the heavy-light current operators come in now at 
$\ord(\alpha_s \, \Lambda_{QCD}/M)$ and $\ord(a \, \alpha_s)$.
The latter two errors are removed by going to the full expression 
in (\ref{a0}).

\section{Summary}
We have performed a one-loop matching of the lattice NRQCD/Asqtad heavy-light 
current (temporal component) to its continuum QCD counterpart, correct 
through $\ord(\alpha_s/M)$ for a range in heavy quark mass.  One-loop 
results for the NRQCD heavy quark self energy corrections and the 
massless Asqtad quark wavefunction renormalization are also presented.
We find that all the perturbative coefficients are well behaved and 
none of them are particularly large.

The matching coefficients are important ingredients in lattice 
investigations of $B$ and $D$ meson leptonic  and semileptonic 
decays. The $\ord(\alpha_s^2)$ errors that remain 
after incorporating the one-loop results of this article will, at some point,
become the dominant error in decay constant and form factor 
determinations from the lattice (this is already happening for
$f_{B_s}$ and $f_{D_s}$ calculations \cite{fbs}). Consequently, 
there is a need to push on to two-loop matching calculations.  
Members of the HPQCD collaboration are already engaged in higher order 
perturbative 
calculations with improved lattice actions \cite{trottier3,quentin1,nobes}. It 
should be possible to extend
those calculations to higher order perturbative matching of 
heavy-light currents as well.

\begin{acknowledgments}
This research is supported by a grant from the US Department of Energy,
 DE-FG02-91ER40690.
  The authors thank Quentin Mason for a copy of his notes on 
Asqtad Feynman rules, Colin Morningstar for sending us 
his $C^{++}$ perturbation theory codes, and Howard Trottier for enabling 
us to use  his codes to check our NRQCD Feynman rules.
\end{acknowledgments}

\appendix

\section{Feynman Rules}

In deriving Feynman rules for fermionic actions we rely heavily on 
methods developed by Colin Morningstar in reference \cite{colin}.
The basic quantity of interest is the ``$\xi$-function'',
\be
\xi_{\hat{A}}^{(r)} =
\xi_{\hat{A}}^{(r)}(\, k^\prime, k \, ; \, (q_1,\nu_1,b_1), (q_2,\nu_2, b_2), 
\ldots, (q_r,\nu_r,b_r) \, )
\ee
which gives the vertex due to operator $\hat{A}$ for a fermion to 
emit $r$ gluons of polarization $\nu_1,\nu_2, \ldots,\nu_r$, color
index $b_1,b_2, \ldots,b_r$ and with momenta $q_1,q_2,\ldots,q_r$. 
$k^\prime$ is the momentum of the outgoing fermion and $k$ the momentum 
of the incoming fermion, $k = k^\prime + \sum_{i=1}^r q_i$.  
Given such vertices for two operators $\hat{A}$ and $\hat{B}$,
one can derive the emission vertices for the product operator $\hat{A} \hat{B}$
using the convolution theorem of Fourier transforms.  The general 
expression is given in eq.(40) of \cite{colin}.  Here we reproduce only the 
special cases of zero- one- and two-gluon emission vertices, which are 
the only ones needed in the one-loop calculations of this article.
\bea
\label{conv0}
\xi_{\hat{A}\hat{B}}^{(0)}(k^\prime,k) &=& \delta_{k^\prime,k} \;
\xi_{\hat{A}}^{(0)}(k,k) \;\; \xi_{\hat{B}}^{(0)}(k,k)  \\
\label{conv1}
\xi_{\hat{A}\hat{B}}^{(1)}(k^\prime,k \, ; \, (q,\nu,b)) &=& 
\xi_{\hat{A}}^{(1)}(k^\prime,k \, ; \, (q,\nu,b)) \;\; \xi_{\hat{B}}^{(0)}(k,k) 
 \; + \nl
&& \xi_{\hat{A}}^{(0)}(k^\prime,k^\prime) \; \;
\xi_{\hat{B}}^{(1)}(k^\prime,k \, ; \, (q,\nu,b))  \\
\label{conv2}
\xi_{\hat{A}\hat{B}}^{(2)}(k^\prime,k \, ; \, (q_1,\nu_1,b_1),(q_2,\nu_2,
b_2)) &=& 
\xi_{\hat{A}}^{(2)}(k^\prime,k \, ; \, (q_1,\nu_1,b_1),(q_2,\nu_2,b_2)) \; \;
\xi_{\hat{B}}^{(0)}(k,k) 
 \; +\nl
    & &     \xi_{\hat{A}}^{(0)}(k^\prime,k^\prime) \; \;
\xi_{\hat{B}}^{(2)}(k^\prime,k \, ; \, (q_1,\nu_1,b_1),(q_2,\nu_2,b_2)) 
 \; + \nl
& & \xi_{\hat{A}}^{(1)}(k^\prime,k^\prime+q_1 \, ; \, (q_1,\nu_1,b_1)) \;
\; \xi_{\hat{B}}^{(1)}(k-q_2,k \, ; \, (q_2,\nu_2,b_2)) \nl
&&
\eea
These equations are extremely useful and also very general. The operators 
$\hat{A}$ or $\hat{B}$ can be simple operators, such as the link 
variable $U_\mu$ or $\nabla_\mu$.  They can also stand for more complicated 
combinations such as $\tilde{F}_{\mu \nu}$ all the way to terms like 
$\left(1 \!-\!\frac{a \delta H}{2}\right)$ or 
$ \left(1\!-\!\frac{aH_0}{2n}\right)^{n}_t 
 \, U^\dagger_t(t-1)
 \left(1\!-\!\frac{aH_0}{2n}\right)^{n}_{t-1} $
 that appear in the NRQCD action.  

\vspace{.1in}
\noindent
Reference \cite{colin} lists the $\xi_{\hat{A}}^{(r)}$ for $r=0,1,2,3$, for a 
wide range of elementary operators, $U_\mu$, $\nabla_\mu$, $\del^{(2)}$, 
$F_{\mu \nu}$ and for improved versions of these operators.  
Using the convolution equations (\ref{conv0}) - (\ref{conv2}), one 
can then build up the emission vertices for composite operators 
such as $\delv\cdot\Ev$ or $(\del^{(2)})^2$ or a string of $U$ matrices. 
This is done, for instance, in the $C^{++}$ code (method 2 of section IID).  
We have also used these equations to obtain Feynman rules 
once and for all by hand.  They 
are summarized in the following two subsections.

\subsection{Feynman Rules for the NRQCD Action}
In this subsection we give the Feynman rules for the NRQCD
action 
 (\ref{nrqcdact}) - (\ref{deltaH}).  We will do so in the two component 
language which is appropriate for self-energy and heavy-heavy current 
matching calculations.  At the end of this subsection we mention the 
trivial generalization to a four component language which is more 
useful for heavy-light current matchings.

\vspace{.1in}
\noindent
In order to streamline the notation and also to introduce spin information,
we re-express the 
relevant $\xi^{(r)}_{\hat{A}}$ as follows.
\be
\label{notation}
 \xi_{\hat{A}}^{(r)}(\, k^\prime, k \, ; \, (q_1,\nu_1,b_1), (q_2,\nu_2, b_2), 
...... (q_r,\nu_r,b_r) \, ) \; \rightarrow \;
  \left[ \hat{A} \right]^{(r)\,\nu_1, ... \nu_r}_s (k^\prime,k) 
\; T^{b_1} .... T^{b_r}
\ee
The gluon momenta $q_1, ...., q_r$ are left implicit in this notation. 
The labels $s = 0,1,2,3$ refer to vertices with spin structure $(I_2, \sigma_x,
\sigma_y, \sigma_z)$ respectively.
We also rewrite the NRQCD action in terms of the following 
operators:
\bea
O_1  &=& \left( 1 - \frac{a \, \delta H }{2} \right)  \\
O_2 &=&  \left(1- \frac{a \, H_0}{2 n} \right)^n  \\
O_{2U2} &=& O_2 \; U_t^\dagger \; O_2
\eea
and schematically one has,
\be
{\cal L}_{NRQCD} = 
\phi^\dagger  \left(\, 1 \; - \; O_1 \, O_{2U2} \, O_1 \, \right)  \phi.
\ee
Feynman rules are obtained by determining,
\be
\left[\, O_1O_{2U2}O_1 \, \right]^{(r)\nu_1,...,\nu_r}_s(k^\prime,k)
\ee
where we are using the notation of (\ref{notation}).

\vspace{.1in}
\noindent
\underline{Tree-level:}  \\
One has
\bea
\left[O_2\right]^{(0)}_s(k^\prime,k) &=&
\delta_{s,0} \, \delta_{k^\prime,k} \; F(k)^n  \\
\left[O_1\right]^{(0)}_s(k^\prime,k) &=& \delta_{s,0} \; \delta_{k^\prime,k}
\; F_1(k) 
\eea
with
\bea
F(k) & = & 
 \left[1 - \frac{1}{n \, (aM_0)}  \sum^3_{j=1} 
 {\rm sin}^2 (\frac{k_j}{2})   \, \right] \\
 F_1(k) &=& 1 - \frac{1}{3} \frac{c_5}{(aM_0)} \sum^3_{j=1} 
 {\rm sin}^4 ( \frac{k_j}{2})  + \left[\frac{c_1}{(aM_0)^3} + 
\frac{c_6}{2n(aM_0)^2} \right] \left[\sum^3_{j=1}  {\rm sin}^2 
( \frac{k_j}{2} )
\right]^2
\eea
and 
\be
\left[O_{2U2}\right]^{(0)}_s(k^\prime,k) =
\delta_{s,0} \, \delta_{k^\prime,k} \; e^{-ik_0} \, F(k)^{2n}
\ee
The free heavy quark propagator is given by,
\bea
G^{(0)}_H(k) &=& \left\{\, 1 \; - \; \left[O_1 O_{2U2} O_1 \right]^{(0)}
_{s=0}(k,k) \, \right\}^{-1} \nl
 &=& \left\{\, 1 \; - \; e^{-ik_0} \, F(k)^{2n} \, F_1(k)^2
 \, \right\}^{-1} \, .
\eea

\vspace{.2in}
\noindent
\underline{One-gluon emission:}  \\
Using the notation of this appendix, the coefficients $wh_{\mu,s}$ of 
eq.(\ref{vh}) are defined (up to a color T-matrix which we omit) as,
\be
wh_{\mu,0}  = 
\left[O_1 O_{2U2} O_1 \right]^{(1)\mu}_{s=0}(k^\prime,k) \quad , \quad
wh_{\mu,j} \, \sigma_j  = 
\left[O_1 O_{2U2} O_1 \right]^{(1)\mu}_{s=j}(k^\prime,k)
\ee
Repeated use of (\ref{conv1}) leads to,
\bea
 \left[O_1 O_{2U2} O_1 \right]^{(1)\mu}_s(k^\prime,k) &=& \;
 \left[O_1 \right]^{(1)\mu}_s(k^\prime,k) \,\left( 
 \left[O_1 O_{2U2} \right]^{(0)}_{s=0}(k^\prime,k^\prime) \; + \;
 \left[ O_{2U2} O_1 \right]^{(0)}_{s=0}(k,k) \right) \; + \nl
 & &  \; 
 \left[O_1 \right]^{(0)}_{s=0}(\kp,\kp) \;
 \left[O_{2U2}\right]^{(1)\mu}_s(k^\prime,k) \;
 \left[O_1 \right]^{(0)}_{s=0}(k,k)  \\
&=&  \;
 \left[O_1 \right]^{(1)\mu}_s(k^\prime,k) \,\left( 
e^{-i \kp_0} \, F(\kp)^{2n} \, F_1(\kp) \; + \;
e^{-i k_0} \, F(k)^{2n} \, F_1(k) \right)  \; + \nl
 && \; F_1(\kp) \, F_1(k) \, \left[O_{2U2}\right]^{(1)\mu}_s(k^\prime,k) \, .
\eea
Going further down the chain one has,
\bea
 \left[O_{2U2}\right]^{(1)\mu=0}_s(k^\prime,k) &=& \delta_{s,0} \;
F(\kp)^n \, F(k)^n \left( -i \, e^{-i(\kp_0 + k_0)/2} \right)  \\
 \left[O_{2U2}\right]^{(1)\mu=j}_s(k^\prime,k) &=& 
 \left[O_2\right]^{(1)j}_s(k^\prime,k) \;
 \left( e^{-i \kp_0} \, F(\kp)^n \; + \; e^{-i k_0} \, F(k)^n \right)  
\eea
with ($ k^\pm \equiv \frac{k \pm \kp}{2}$) 
\be
 \left[O_2\right]^{(1)j}_s(k^\prime,k)  = 
\delta_{s,0} \; \frac{-1}{2n (aM_0)} \; {\rm sin}(k^+_j) \; 
F_2(\kp,k,n) 
\ee
\be
F_2(\kp,k,n)  = \left\{ \begin{array}{l}
                1 \qquad \qquad \qquad \qquad \qquad \qquad n=1  \\
               \sum_{l=0}^{n-1} F(\kp)^{n-l-1} \; F(k)^l  \quad \qquad n > 1
          \end{array} \right.   
\ee
and
\be
 \left[O_1\right]^{(1)\mu}_s(k^\prime,k) = - \; \frac{a}{2}
\sum_{i=1}^6 
 \left[\delta H_i \right]^{(1)\mu}_s(\kp,k) \, .
\ee
The 
$ \left[\delta H_i \right]^{(1)\mu}_s(\kp,k) $ can be built up 
from the Fourier transforms listed in \cite{colin} using again 
(\ref{conv1}). One finds,
\bea
 && - \; \frac{a}{2}  \;
 \left[\delta H_1 \, + \, \delta H_6 \right]^{(1)j}_s(\kp,k)  = \nl
 && \delta_{s,0} \; \left( \frac{c_1}{2 (aM_0)^3} \, + \, \frac{c_6}
{4n(aM_0)^2} \right) \; {\rm sin}(k^+_j) \; \left[
 \sum^3_{l=1}  {\rm sin}^2 (\frac{k_l}{2} )  \, + \,
 \sum^3_{l=1}  {\rm sin}^2 ( \frac{\kp_l}{2} )  \, 
\right] 
\eea
\vspace{.1in}
\bea
 - \; \frac{a}{2}  \;
 \left[\delta H_2 \right]^{(1)0}_s(\kp,k) &=& \delta_{s,0} 
\,
\frac{-i c_2}{16 (aM_0)^2} \, \sum^3_{l=1} 
\left\{ {\rm sin}(2k^-_l) \, {\rm cos}(k^-_0) \, \left[ {\rm sin}(\kp_l) - {\rm sin}(k_l)
\right] \, \eta_{l0} \right\}  \nl
   &&  \\
 - \; \frac{a}{2}  \;
 \left[\delta H_2 \right]^{(1)j}_s(\kp,k) &=& \delta_{s,0} 
\,
\frac{i c_2}{16 (aM_0)^2} \, 
\left\{ {\rm sin}(2k^-_0) \, {\rm cos}(k^-_j) \, \left[ {\rm sin}(\kp_j) - {\rm sin}(k_j)
\right] \, \eta_{j0} \right\}  \nl
&&  
\eea
\vspace{.1in}
\bea
 - \; \frac{a}{2}  \;
 \left[\delta H_3 \right]^{(1)0}_{s=i}(\kp,k) &=& \frac{c_3}{16 (aM_0)^2}
\; \sigma_i \; \sum^3_{j,l=1} 
\epsilon_{ijl}\;{\rm cos}(k^-_0)  \left[ ss(\kp_j) \, + \, ss(k_j)
 \right] \,{\rm sin}(2k^-_l) \, \eta_{l0}  \nl
  && \\
 - \; \frac{a}{2}  \;
 \left[\delta H_3 \right]^{(1)j}_{s=i}(\kp,k) &=& \frac{c_3}{16 (aM_0)^2}
\; \sigma_i \; \sum^3_{l=1}
 \epsilon_{ijl}\;{\rm cos}(k^-_j)  \left[ ss(\kp_l) \, + \, ss(k_l)
 \right] \,{\rm sin}(2k^-_0) \, \eta_{j0}   \nl
&& 
\eea
\vspace{.1in}
\be
 - \; \frac{a}{2}  \;
 \left[\delta H_4 \right]^{(1)j}_{s=i}(\kp,k) ~=~ \frac{- i c_4}{4 (aM_0)}
\; \sigma_i \; \sum^3_{l=1}
 \epsilon_{ijl}\;{\rm cos}(k^-_j)  \,{\rm sin}(2k^-_l) \, \eta_{jl}   
\ee
\vspace{.1in}
\be
\label{deltah5}
 - \; \frac{a}{2}  \;
 \left[\delta H_5 \right]^{(1)j}_s(\kp,k) ~=~ \delta_{s,0} 
\; \frac{c_5}{6(aM_0)} \; \left[ - \, {\rm sin}(k^+_j) \; + \; \frac{1}{2} 
\,{\rm sin}(2k^+_j) \,{\rm cos}(k^-_j) \right]
\ee
where $\eta_{\mu \nu} = 1$ for unimproved ${\bf E}$ and ${\bf B}$ fields 
and 
$ \eta_{\mu \nu} = \frac{1}{3}[5 - {\rm cos}(2k^-_\mu) - 
{\rm cos}(2 k^-_\nu)]$ 
for improved fields.  The function $ss(k_j) = {\rm sin}(k_j)$ for 
unimproved $\nabla_j$ and $ss(k_j) = \frac{4}{3} {\rm sin}(k_j) -
\frac{1}{6} {\rm sin}(2k_j)$ for the improved $\tilde{\nabla}_j$.

\vspace{.1in}
\noindent
We have coded up the one-gluon emission vertices, 
$ \left[O_{2U2}\right]^{(1)\mu}_s(k^\prime,k)$, 
$ \left[O_1 \right]^{(1)\mu}_s(k^\prime,k)$, etc.\ as functions of 
arbitrary incoming and outgoing momenta, $k$ and $\kp$.  These routines 
are called repeatedly by the VEGAS code when calculating the integrand 
of the one-loop diagrams. Similar expressions are also fed into 
Mathematica routines, whenever one needs to take derivatives 
with respect to external momenta.

\vspace{.2in}
\noindent
\underline{Two-gluon emission:}  \\
We will restrict the discussion to two-gluon emission relevant for 
the tadpole diagram, for which only a special combination of 
momenta are required : $(\kp,k,q_1,q_2) \rightarrow (k,k,q,-q)$. 
One is also interested only in the spin singlet contribution. 
Applying (\ref{conv2}) to this case one has,
\bea
\left[\, O_1O_{2U2}O_1 \, \right]^{(2)\mu,\nu}_{s=0}(k,k)
 &=& 
\left[O_1 \right]^{(2)\mu,\nu}_{s=0}(k,k) \;  2 \, e^{-ik_0} \,
F(k)^{2n} \, F_1(k)  \; + \; 
\left[O_{2U2} \right]^{(2)\mu,\nu}_{s=0}(k,k) \;  F_1(k)^2 \; + \; \nl
&& F_1(k) \, 
\left[O_{2U2} \right]^{(1)\mu}_{s=0}(k,k+q) \,
\left[O_1 \right]^{(1)\nu}_{s=0}(k+q,k) \; + \nl
&& \left[O_1 \right]^{(1)\mu}_{s=0}(k,k+q) \,
\left[O_{2U2} \right]^{(1)\nu}_{s=0}(k+q,k) \, F_1(k)  \; + \; \nl
&&  
\left( \left[O_1 \right]^{(1)\mu}_{s_1}(k,k+q) \,
\left[O_1 \right]^{(1)\nu}_{s_2}(k+q,k) \right)\bigg|_{s=0} 
 \; e^{-i(k_0+q_0)} \;F(k+q)^{2n} \, . \nl
&&
\eea
The two new ingredients are 
$ \left[O_1 \right]^{(2)\mu ,\nu}_{s=0}(k,k,q,-q) $ and 
$\left[O_{2U2} \right]^{(2)\mu,\nu}_{s=0}(k,k,q,-q)$.  
\bea
 \left[O_1 \right]^{(2)\mu ,\nu}_{s=0}(k,k,q,-q) &=&
\left( \frac{c_1}{2 (aM_0)^3} \, + \, \frac{c_6}{4n(aM_0)^2} \right)
\; 
\delta_{\mu,i} \, \delta_{\nu,j} \nl
&&  \left[ \delta_{ij} \, {\rm cos}(k_j) 
\, \sum^3_{l=1}  {\rm sin}^2 (\frac{k_l}{2} )
 \; + \frac{1}{2}
{\rm sin}(k_i + \frac{q_i}{2}) \; {\rm sin}(k_j + \frac{q_j}{2}) \right]  \nl
&+& \frac{ i c_2}{16 (aM_0)^2} \bigg[ (\delta_{\mu,j} \delta_{\nu,0} 
\, + \delta_{\mu,0} \delta_{\nu,j} ) 
\, {\rm cos}(k_j+\frac{q_j}{2}) \, {\rm sin}(q_j) \,
 {\rm cos}(\frac{q_0}{2}) \, \eta_{j0}  \nl
&&  \qquad \qquad \qquad - \; \delta_{\mu,j} \delta_{\nu,j} \,
2 \, {\rm cos}(k_j+\frac{q_j}{2}) \, 
{\rm sin}(q_0) \, {\rm cos}(\frac{q_j}{2}) \, \eta_{j0}  \bigg]  \nl
&+&  \frac{-c_5}{12 (aM_0)} \; \delta_{\mu,j} \delta_{\nu,j} \; 
 \left[ {\rm cos}(k_j) \, - \, {\rm cos}(2k_j) \,
 {\rm cos}^2(\frac{q_j}{2})  \right]
\eea
where $\eta_{j0}$ is the same as the $\eta_{\mu \nu}$ 
defined  after 
eq.(\ref{deltah5}) with $2 k^-_\mu \rightarrow q_\mu$.
\bea
 \left[O_{2U2} \right]^{(2)\mu ,\nu}_{s=0}(k,k,q,-q) &=&
2 \, e^{-ik_0} \, F(k)^n \, 
 \left[O_2 \right]^{(2)\mu ,\nu}_{s=0}(k,k,q,-q)  \; - \; 
\delta_{\mu,0} \delta_{\nu,0} \,\frac{1}{2} \,
e^{-ik_0} \, F(k)^{2n} \nl
& - & \, i \, F(k)^n \, e^{-i(k_0+\frac{q_0}{2})} \, \left[ 
\delta_{\mu,0} \, \left[O_2 \right]^{(1)\nu}_{s=0}(k+q,k)  \; + \; 
\delta_{\nu,0} \, \left[O_2 \right]^{(1)\mu}_{s=0}(k,k+q)  \right] \nl
& + & \; e^{-i(k_0 + q_0)} \,
 \left[O_2 \right]^{(1)\mu}_{s=0}(k,k+q) \,
\left[O_2 \right]^{(1)\nu}_{s=0}(k+q,k) 
\eea
where
\bea
 \left[O_2 \right]^{(2)\mu \nu}_{s=0}(k,k,q,-q) &=&
 \; \delta_{\mu,i} \delta_{\nu,j} \;  \left[ 
\delta_{ij} \, n \, F(k)^{n-1} 
\, \frac{-1}{4n (aM_0)} \, {\rm cos}(k_j) \right. \; + \nl
&& \left. \qquad  \left( \frac{1}{2n(aM_0)}
\right)^2 \, {\rm sin}(k_i+\frac{q_i}{2}) \,
 {\rm sin}(k_j+\frac{q_j}{2}) \, F_3(k,k+q,n) \right]
\nl
&&
\eea
with
\be
F_3(k,k+q,n)  = \left\{ \begin{array}{l}
         0 \qquad \qquad \qquad \qquad \qquad \qquad \qquad n=1  \\
         \sum_{l=1}^{n-1} l \,
 F(k)^{l-1} \; F(k+q)^{n-l-1}  \quad \quad n > 1
          \end{array} \right. \, .
\ee

\vspace{.2in}
\noindent
\underline{4 component formulas :} \\
So far we have written down Feynman rules corresponding to  
the NRQCD action (\ref{nrqcdact}) which is given in terms of 
2 component Pauli spinors.  One can equivalently write rules 
appropriate for the $Q$ fields of (\ref{j0}) - (\ref{j2}). 
One needs to make the replacement 
\be
(I_2,{\bf \sigma}) \; \rightarrow \; (I_4, {\bf \Sigma})
\ee
with $\Sigma_i = {\rm diag}(\sigma_i,\sigma_i)$ denoting $4 \times 4$ 
matrices.  Furthermore,
 the above gluon  emission vertices should be 
multiplied by $\frac{I + \gamma_0}{2}$.

\subsection{Feynman Rules for the Asqtad Action}
As mentioned already in the main text, Feynman rules for 
the Asqtad action for one-gluon emission and the most general two-gluon 
emission vertices have been written down by Q. Mason \cite{quentin}. 
 In this appendix 
we list just those rules used in our calculations.  

\vspace{.2in}
\noindent
\underline{Tree-level :}  \\
The free Asqtad propagator is given by,
\be
G^{(0)}_q(k)  = \left\{ \, i \, \sum_\mu \gamma_\mu \, {\rm sin}(k_\mu) \, \left[
 1 \, + \, \frac{1}{6} \, {\rm sin}^2(k_\mu) \right] \; + \; (am) \, 
\right\}^{-1} \, .
\ee

\vspace{.2in}
\noindent
\underline{One-gluon emission :}  \\
Carrying out the derivatives and symmetrizations of eq.(\ref{vmu}) 
one ends up with a fattened link consisting of [1-link, 3-link, 
nonplanar 5-link, nonplanar 7-link] staples with weights 
[1/8, 1/16, 1/64, 1/384] respectively \cite{doug}.
  The Lepage term leads to 
a 1-link contribution to $V^\prime_\mu$ with weight 3/8 and a planar 
5-link contribution with weight -1/16.  
Adding to this the Naik term one ends up with,
\be
\left[{\rm Asqtad} \right]^{(1) \mu} = 
\left[ {\rm fat \; link} \right]^{(1) \mu} \; + \;
\left[ {\rm Lepage} \right]^{(1) \mu} \; + \;
\left[{\rm  Naik} \right]^{(1) \mu} 
\ee
with (we again define $ k^\pm \equiv \frac{k \pm \kp}{2}$) 
\bea
\left[ {\rm fat \; link} \right]^{(1) \mu}(\kp,k) &=&  - \, i \, \left\{
 \, \gamma_\mu \, {\rm cos}(k^+_\mu)
 \, {\rm cos}^2(k^-_\nu) 
 \, {\rm cos}^2(k^-_\rho) 
 \, {\rm cos}^2(k^-_\sigma) 
 \right.  \nl
&+&  \sum_{\nu \neq \mu} \gamma_\nu \, {\rm cos}(k^+_\nu) \, 
{\rm sin}(k^-_\mu) \, {\rm sin}(k^-_\nu) \, 
\left[ \frac{1}{3} \, + \, \frac{1}{6} \, \left( {\rm cos}^2(k^-_\sigma) 
+ {\rm cos}^2(k^-_\rho) \right) \right.  \nl
&& \left. \left.\qquad \qquad + \, \frac{1}{3} \, {\rm cos}^2(k^-_\sigma) 
\, {\rm cos}^2(k^-_\rho) \right] \; \right\}   \\
\left[ {\rm Lepage} \right]^{(1) \mu}(\kp,k) &=&  - \, i \, \left\{
 \, \gamma_\mu \, {\rm cos}(k^+_\mu) \, \frac{1}{4} \, \sum_{\nu \neq \mu} 
{\rm sin}^2(2k^-_\nu) \right.  \nl
&&  \qquad \qquad - \;  \left.
 \sum_{\nu \neq \mu} \gamma_\nu \, {\rm cos}(k^+_\nu) \, 
{\rm sin}(k^-_\mu) \, {\rm sin}(k^-_\nu) \, {\rm cos}^2(k^-_\mu) \right\} \\
\left[ {\rm Naik} \right]^{(1) \mu}(\kp,k) &=&  - \, i \, \left\{
 \, \gamma_\mu \, {\rm cos}(k^+_\mu) \, \frac{1}{6} \,
\left[ {\rm cos}^2(k^+_\mu) \, - \, {\rm cos}^2(k^-_\mu) 
\, \left( 1 \, - \, 4 \, {\rm sin}^2(k^+_\mu) \right) \right ] \right\}\, .\nl
&&
\eea
In the above formulas, the Lorentz indices $\mu,  \; \nu, \; \rho, \; 
\sigma$ are all taken to be different from each other.
One can read off from these equations the coefficients $w_{\mu,\nu}$ 
of eq.(\ref{vq}).

\vspace{.2in}
\noindent
\underline{Two-gluon emission :}  \\
We will again present only those two-gluon emission vertices 
specific to the tadpole diagram where $(\kp,k,q_1,q_2) \rightarrow 
(k,k,q,-q)$. One has
\bea
\left[{\rm Asqtad} \right]^{(2) \mu,\mu} &=& 
\frac{i}{2} \, \left\{ \gamma_\mu \, {\rm sin}(k_\mu) \, \left[ 1 \; + \; 
\frac{1}{24} \, \left( [1-4 {\rm cos}^2(k_\mu) ][ 1 
- 4{\rm cos}^2(\frac{q_\mu}{2})]^2 
\, + \, 3 \right) \right] \right. \nl
&& \left. \qquad + \; \sum_{\nu \neq \mu} \gamma_\nu \, {\rm sin}(k_\nu) \,\, 
{\rm sin}^2(\frac{q_\nu}{2}) \, 2 \, {\rm sin}^2(\frac{q_\mu}{2}) \right\}  \\
\left[{\rm Asqtad} \right]^{(2) \mu \neq \nu} &=& 
\frac{i}{2} \, \left\{ \gamma_\mu \, {\rm sin}(k_\mu) \, 
\left[ \, - \,\frac{1}{6} \, 
{\rm sin}(\frac{q_\mu}{2}) \, {\rm sin}(\frac{q_\nu}{2})
 \left( 2 \, + \, {\rm cos}^2(\frac{q_\rho}{2}) \, + \, 
{\rm cos}^2(\frac{q_\sigma}{2}) \right. \right. \right. \nl
&& \left. \left. 
 \qquad \, + \, 2 \, {\rm cos}^2(\frac{q_\rho}{2}) \, {\rm cos}^2(\frac{q_\sigma}{2})
 \right) \; + \; \frac{1}{2} \, {\rm sin}(\frac{q_\mu}{2}) 
\, {\rm cos}(\frac{q_\nu}{2}) \, {\rm sin}(q_\nu) \right]  \nl
&-& 
 \; \sum_{\rho \neq \mu , \nu} \gamma_\rho \, 
{\rm sin}(k_\rho) 
\, {\rm sin}^2(\frac{q_\rho}{2}) 
\, \frac{1}{6} 
\, {\rm sin}(\frac{q_\mu}{2}) 
\, {\rm sin}(\frac{q_\nu}{2}) 
\, \left( 2 \, + \, {\rm cos}^2(\frac{q_\sigma}{2}) \right)  \nl
&&  \qquad \qquad + \qquad [ \; \mu \; \rightleftharpoons \; \nu 
\; ] \quad \bigg\} \, .
\eea

\section{Infrared Subtractions}
At intermediate stages of the matching procedure 
several of the lattice one-loop integrals are IR divergent.
 We use a gluon mass 
$\lambda$ to regulate those integrals and extract the IR finite 
contributions in the following way (similar approaches can be 
found, for instance, in \cite{kura}\cite{groote}).
 If ${\cal F}_L(k,\lambda,M_0)$ is 
the relevant lattice integrand, one can write,

\be
\label{zsub}
  \int_k {\cal F}_L(k, \lambda,M_0) \;  = \;
  \int_k 
\left[ \, {\cal F}_L(k,\lambda,M_0) 
 \; - \; {\cal F}_{sub}(k,\Lambda,\lambda,M_0) \,  \right ]
 \quad + \quad R(\Lambda,\lambda,M_0),
\ee
with
\be
\label{addback}
 R(\Lambda,\lambda,M_0) = 
 \int_k  {\cal F}_{sub}(k,\Lambda,\lambda,M_0) 
\ee
and
\be
\int_k \equiv g^2 \frac{4}{3}\int\frac{d^4k}{(2 \pi)^4} \, .
\ee
To be consistent with the previous Appendix,  the momentum $k$ is always 
measured in units of the inverse lattice spacing. 
 $\Lambda$, $\lambda$ and $M_0$ however, have dimensions of 
energy.
${\cal F}_{sub}(k,\Lambda,\lambda,M_0)$ is constructed to have the 
same IR behavior as the lattice integrand of interest, so that 
the first integral on the RHS of (\ref{zsub}) is IR finite.
$\Lambda$ is a (ultraviolet) cutoff imposed on the
subtraction term so that ${\cal F}_{sub}(k,\Lambda,\lambda,M_0) = 0$ 
for $k^2 > a^2 \,\Lambda^2$.  We take $\Lambda \leq \frac{\pi}{a}$ 
and vary $\Lambda$ to 
check that the dependence on this cutoff cancels between the two terms
in (\ref{zsub}).  Another criterion for choosing a suitable ${\cal F}_{sub}$ 
is that the integral in (\ref{addback}) be easy to do and that one be able to 
extract the IR divergent piece, $R^{IR}$, analytically:
\be
R(\Lambda,\lambda,M_0) = R^{finite}(\Lambda,M_0) \; + \; 
R^{IR}(\Lambda,\lambda,M_0) \, .
\ee
Otherwise there is a lot of freedom in choosing ${\cal F}_{sub}$.
We give one example for the logarithmic IR divergence in 
the NRQCD wavefunction renormalization $Z_Q$

\vspace{.1in}
\noindent
The IR divergence in $Z_Q$ comes from $\Omega_1$ defined in (\ref{omega1}).
A convenient choice for ${\cal F}_{sub}$ in this case is the corresponding 
quantity in continuum NRQCD, which must have the same IR behavior as the 
lattice theory.  One can easily convince oneself that the full continuum 
${\bf \Sigma}$ is actually not required. The IR divergence resides in 
the contribution to ${\bf \Sigma}$ from a temporal gluon in the 
one-loop rainbow diagram of Fig.~1, the only contribution that survives 
into the static limit. 
Setting $\mu = \nu = 0$ in 
the gluon propagator one finds,
\be
\label{omega1sub}
{\cal F}^{\Omega1}_{sub}(k,\Lambda,\lambda,M_0)   = \left\{ \begin{array}{l}
                 \frac{w^2 \, - \, k_0^2 }{[ k_0^2 + w^2]^2}
\; \frac{1}{\left( \,k^2 + (a \lambda)^2 \, \right)} 
\; \left [ \, 1 \, + \, (\xi - 1) \frac{k_0^2}{k^2} \right]  
 \qquad k^2 \, \leq \, a^2 \Lambda^2 \\
                                                             \\
                0 \qquad \qquad \qquad \qquad \qquad \qquad \qquad  \qquad
                                     \; \;   k^2 \,> \, a^2 \Lambda^2  \\
          \end{array} \right.   
\ee
with,
\be
w \equiv \frac{\vec{k}^2}{2 \, (aM_0)} \, .
\ee
Using 4D spherical coordinates the angular integrals in (\ref{addback})
can be carried out analytically (we use MAPLE for this)
and one is left with a 1D radial integral,
\bea
R(\Lambda,\lambda,M_0) &=& 
\frac{g^2}{4 \pi} \; \frac{1}{3  \pi} \, \int_0^{a^2\Lambda^2}  
\, \frac{dy}{(y + (a \lambda)^2)} \; \left \{ \frac {2}
{\left[ 2 \sqrt{y b } \, + \, 1 \right] ^{3/2}} \right.
 \; +  \nonumber \\
 & & \left.\; 2 \, ( \xi - 1) \; \frac { yb + 3 \sqrt{y b} + 1 - 
\left[ 2 \sqrt{y b } \, + \, 1 \right] ^{3/2} } {
y b \, \left[ 2 \sqrt{y b } \, + \, 1 \right] ^{3/2} } \right \}
\eea
where,
\be
b \equiv \frac{1}{4 \, (aM_0)^2} \, .
\ee
The IR divergent contribution to $R(\Lambda,\lambda,M_0)$ can be isolated 
straightforwardly and one has,
\bea
R^{finite} &=& 
 \frac{\alpha_s}{3  \pi} \; \left( \; \int_0^{a^2\Lambda^2}  
\, \frac{dy}{(y + (a \lambda)^2)} \; \left \{ \left[ \frac {2}
{\left[ 2 \sqrt{y b } \, + \, 1 \right] ^{3/2}} \; - \; 2 \right] \right.
 \right . \; +  \nonumber \\
 & & 
\left. \left. \;  ( \xi - 1) \; 
\left[ \, 2 \; \frac{ yb + 3 \sqrt{y b} + 1 - 
\left[ 2 \sqrt{y b } \, + \, 1\right] ^{3/2} } {
y b \, \left[ 2 \sqrt{y b } \, + \, 1 \right] ^{3/2} } \; + \; 1 \,
\right]  \right\} \; \right)_{a \lambda \, \rightarrow \, 0}
 \\
R^{IR} &=& \frac{\alpha_s}{3 \pi} \; \int_0^{a^2\Lambda^2} \, 
\frac{dy}{(y + (a \lambda)^2)} \;  [ 2 - (\xi - 1)  ]  \nonumber \\
 &=& \frac{\alpha_s}{3 \pi} \;
[ 2 - (\xi - 1) ] \; {\rm ln}\frac{\Lambda^2}{\lambda^2}  \, .
\eea
The $\Lambda$ dependence is canceled by the first term on the RHS 
of (\ref{zsub}). The gluon mass dependence is the same as in continuum 
QCD.

\newpage

%\begin{references}

%\end{references}

%%%%%%%%%%%%%%%%%%%%%%%%%%%%%%%%%%%%%%%%%%%%%%%%%%%%%%%%%%%%%%%%%%%%%%
%
%                       TABLES
%
%%%%%%%%%%%%%%%%%%%%%%%%%%%%%%%%%%%%%%%%%%%%%%%%%%%%%%%%%%%%%%%%%%%%%

\begin{table}
\caption{Coefficient of the one-loop contribution to $Z_q$ for 
massless Asqtad quarks.  The plaquette definition
of $u_0$ is used to implement tadpole improvement. $\xi$ is the gauge 
parameter.
 Where no errors are indicated, they are of order one or less in the 
last digit.
  }
\begin{center}
\begin{tabular}{|c|ccc|c|}
\hline
$\quad \xi \quad$    &  $\qquad C_q^{reg} \qquad$ 
 &  $\qquad C_q^{tad} \qquad$  &
  $\qquad C_q^{u0} \qquad$  & $ \quad \qquad C_q \qquad$  \\
\hline
 1       &  $-0.798(3)$  &  $ 1.600 $    &  $-1.726$   & $-0.924(3)$  \\
 0       &  $\;\;0.000(3)$& $ 1.310 $    &  $-1.726$   & $-0.416(3)$  \\
\hline
\end{tabular}
\end{center}
\end{table}

\begin{table}
\caption{Coefficient of one-loop contributions to the heavy quark 
self-energy.
  The Landau link definition
of $u_0$ is used to implement tadpole improvement. $\xi$ is the gauge 
parameter. 
 Where no errors are indicated, they are of order one or less in the 
last digit.
  }
\begin{center}
\begin{tabular}{|cc|cccc|}
\hline
$\quad aM_0 \quad$ &  $ \quad n \quad $   &  $\quad C_Q(\xi = 1) \quad$ 
 &  $\quad C_Q(\xi = 0)\quad$  &
  $\qquad C_M \qquad$  & $  \qquad C_{E0} \qquad$  \\
\hline
 5.40   &  1 & $-0.016(3)$ & $\;\;0.493(3)$ & $-0.026(4)$ & $0.917$ \\
 4.00   &  2 & $-0.142(3)$ & $\;\;0.368(3)$ & $\;\;0.082(4)$ & $0.850$\\
 2.80   &  2 & $-0.338(3)$ & $\;\;0.171(3)$ & $\;\;0.235(4)$ & $0.767$\\
 2.10   &  4 & $-0.539(3)$ & $-0.030(3)$ & $\;\;0.387(4)$ & $0.666$\\
 1.95   &  2 & $-0.611(3)$ & $-0.105(3)$ & $\;\;0.421(4)$ & $0.689$\\
 1.95   &  4 & $-0.603(3)$ & $-0.094(3)$ & $\;\;0.428(4)$ & $0.646$\\
 1.60   &  4 & $-0.797(3)$ & $-0.288(3)$ & $\;\;0.543(4)$ & $0.609$\\
 1.20   &  6 & $-1.184(3)$ & $-0.678(3)$ & $\;\;0.732(4)$ & $0.609$\\
 1.00   &  6 & $-1.545(3)$ & $-1.037(3)$ & $\;\;0.859(4)$ & $0.758$\\
\hline
\end{tabular}
\end{center}
\end{table}

\begin{table}
\caption{ Matching Coefficients for the temporal components 
of the heavy-light axial  and vector currents. The Landau link $u_0$ 
was used to evaluate $\rho_1$. 
 Where no errors are indicated, they are of order one or less in the 
last digit.
  }
\begin{center}
\begin{tabular}{|cc|cccc|}
\hline
$\quad aM_0 \quad$ &  $ \quad n \quad $   &  $\quad \quad \tilde{\rho}_0
\quad \quad$ 
 &  $\quad \quad \rho_1 \quad \quad$  &
  $\quad \quad \rho_2 \quad \quad$  & $  \quad \quad \zeta_{10}
\quad \quad$  \\
\hline
 5.40   &  1 & $\;\;0.240(2)$ & $0.509(4)$ & $-0.155(6)$& $-0.095$\\
 4.00   &  2 & $\;\;0.149(2)$ & $0.445(4)$ & $-0.134(6)$& $-0.123$\\
 2.80   &  2 & $\;\;0.043(2)$ & $0.382(4)$ & $-0.072(6)$& $-0.166$ \\
 2.10   &  4 & $-0.044(2)$ & $0.346(4)$ & $\;\;0.002(6)$& $-0.210$\\
 1.95   &  2 & $-0.058(3)$ & $0.343(4)$ & $\;\;0.008(6)$& $-0.222$\\
 1.95   &  4 & $-0.065(3)$ & $0.342(4)$ & $\;\;0.018(6)$& $-0.219$\\
 1.60   &  4 & $-0.117(2)$ & $0.343(4)$ & $\;\;0.061(6)$& $-0.259$\\
 1.20   &  6 & $-0.175(2)$ & $0.385(4)$ & $\;\;0.107(6)$& $-0.329$ \\
 1.00   &  6 & $-0.186(2)$ & $0.463(4)$ & $\;\;0.119(6)$& $-0.378$ \\
\hline
\end{tabular}
\end{center}
\end{table}

\end{document}